# The Solar Argon Abundance

Katharina Lodders[1]

**Abstract**: The solar argon abundance cannot be directly derived by spectroscopic observations of the solar photosphere. The solar Ar abundance is evaluated from solar wind measurements, nucleosynthetic arguments, observations of B stars, HII regions, planetary nebulae, and noble gas abundances measured in Jupiter's atmosphere. These data lead to a recommended argon abundance of $N(Ar) = 91{,}200 \pm 23{,}700$ (on a scale where $Si = 10^6$ atoms). The recommended abundance for the solar photosphere (on a scale where $\log N(H) = 12$) is $A(Ar)_{photo} = 6.50 \pm 0.10$, and taking element settling into account, the solar system (protosolar) abundance is $A(Ar)_{solsys} = 6.57 \pm 0.10$.

Subject headings: Sun: abundances – photosphere-corona-abundances

[1] Planetary Chemistry Laboratory, Department of Earth and Planetary Sciences and McDonnell Center for the Space Sciences, Washington University, Campus Box 1169, St. Louis, MO63130-4899; lodders@levee.wustl.edu

## 1. Introduction

The solar Ar abundance quoted in two recent compilations varies by a factor of 2.3. On the logarithmic abundance scale fixed to $A(H) = \log N(H) = 12$, the selected photospheric Ar abundance by Lodders (2003) is $A(Ar) = 6.55\pm0.08$, and Asplund et al. (2005) give $A(Ar) = 6.18\pm0.08$. The Ar abundance cannot be determined spectroscopically in the solar photosphere and indirect methods have to be employed to derive its abundance. Lodders (2003) based the Ar abundance on nucleosynthesis arguments (see §4) whereas Asplund et al. (2005) based the Ar abundance on solar energetic particle measurements (see §2). The use of two different methods seems to be the reason for the difference in the quoted solar Ar abundance in these compilations.

However, such a relatively large difference is unsatisfactory for an elemental abundance that serves as reference standard. For example, the abundance of Ar and the other noble gases is quite important for modeling the evolution of the terrestrial and Venusian atmospheres (e.g., Lewis & Prinn, 1984). In addition to the constraints from nucleosynthesis and the solar wind, the Ar abundances in B stars, HII regions, planetary nebulae, and in Jupiter's atmosphere can be used to estimate the solar Ar abundance. The different constraints are described next and summarized in Table 1 and Figure 1.

## 2. Argon abundance from solar corpuscular radiation

The solar argon abundance can be derived from abundance ratios of Ar to other elements measured in the solar wind (SW) and solar energetic particles (SEP) coming from the Sun's corona. If the photospheric abundance of the reference element in the Ar/element ratio is well known, the Ar abundance follows from the ratio. The limitation of this method is that elements may become fractionated in the SW and SEP from photospheric abundances according to their different first ionization potentials (FIPs). Relative to photospheric values, elements with low FIP < 10eV (e.g., Mg, Ca) appear to be enriched in the SW and SEP, whereas elements with high FIP (e.g., N, O, Ne, Ar) seem to have retained relative photospheric abundances (see discussions by Meyer 1993 and Feldman & Laming 2000). These conclusions are based on SW and SEP abundances relative to older solar reference abundances (e.g., Anders & Grevesse (1989) or Grevesse & Sauval (1998)). The solar abundances of all high FIP elements relevant here (N, O, Ne) and the intermediate FIP element C, which were used to define the size of the FIP fractionations, have been revised downwards (Lodders 2003, Asplund et al. 2005). Thus, the relative fractionation factors between the high and low FIP elements should be re-evaluated for the different SW and SEP components.

Some of the Ar abundances below are from Ar/O ratios, so the Ar abundances depend on the adopted solar O abundance. Here I use the new O-abundance from Allende-Prieto et al. (2001) to keep the calculations consistent with the abundances in Lodders (2003). The new lower solar O abundance agrees better with the O abundance in the local ISM, which appeared to be too low when compared to the older, larger O abundance, and is supported by measurements in B stars (e.g., Cunha et al. 2006). Asplund et al. (2005) give a slightly lower O abundance (0.03 dex less), and if this O abundance is preferred, the Ar abundance from the Ar/O ratio changes accordingly.

**2.1. Argon abundance from solar wind**

Measurements of the solar wind captured in meteoritic iron-nickel give $N(^4He)/N(^{36}Ar) = 30,400\pm2,200$ (Murer et al. 1997). Assuming that He and Ar are not fractionated from each other, and using the photospheric He abundance $N(He) = 2.288\times10^9$ (on a scale where $N(Si) = 10^6$ atoms; Lodders 2003), gives $N(^{36}Ar) = 75,300\pm5,500$. The fraction of $^{36}Ar$ in total elemental Ar is 84.6%, hence $N(Ar) = 89,000\pm6,400$, or, on a scale where $\log N(H) = A(H) = 12$, we obtain $A(Ar) = \log N(Ar) + 1.54 = 6.49\pm0.03$ for the photosphere (see Lodders 2003 for how the different abundance scales are linked). Helium has the highest FIP (24.6 eV) of any element and it may fractionate from other high FIP elements (see Meyer 1993, Feldman & Laming 2000). Murer et al. (1997) only give data for He, Ne, and Ar. Measurements by the Solar Wind Ion Composition Spectrometer on Ulysses (Steiger et al. 2000) show that (relative to the newer solar N/Mg and O/Mg ratios; Lodders [2003]), the abundances of N and O are ~0.4 times solar in the fast SW, and ~0.6 times solar in the slow SW. The He/Mg ratio in the slow and fast solar wind is ~0.3 times solar. Steiger et al. (2000) do not give data for Ar, but if Ar behaves like O and N, this indicates that Ar/He ratios in the solar wind may range from about solar to up to two times solar. In that case, the Ar abundance from the Ar/He ratio by Murer et al. (1997) only is an upper limit. In principle, the fractionation could be constrained from the Ar/Ne ratio. However, the solar Ne abundance has considerable uncertainty and is under debate (e.g., Bahcall et al. 2005) and Ne is not used here to constrain the Ar abundance. Considering the overall uncertainties, the Ar abundance from the solar wind in meteoritic metal is A(Ar) $\leq$6.49($\pm$0.03).

One could also derive the Ar abundance from SW measurements of the heavy noble gases Ar, Kr, and Xe captured in lunar soil or meteorite samples (Wieler & Baur, 1995, Wieler 1998). However, the Ar/Kr and Kr/Xe ratios in the lunar soil samples show variations that appear to correlate with sample irradiation age. In addition, fractionations related to FIP (or first ionization time) are present (see Wieler & Baur 1995, Wieler 1998, Heber et al. 2001). Because of these complications, these data are not used to constrain the solar Ar abundance.

Young et al. (1997) re-evaluated the photospheric Ar abundance as 6.47$\pm$0.10, based mainly on the N(Ar)/N(O) ratio of 0.004$\pm$0.001 for the solar wind in lunar soils. They re-evaluated the Ar abundance in order to compare the photospheric Ar/Ca ratio to their Ar/Ca measurements of coronal plasma, where Ar and Ca fractionated because of the FIP effect. Young et al. (1997) adopted A(O) = 8.87$\pm$0.07 to calculate their Ar abundance. The more recent photospheric abundance of A(O)=8.69$\pm$0.05 from Allende-Prieto et al. (2001) gives A(Ar) = 6.29$\pm$0.10 or N(Ar) = 56,000$\pm$14,000 from the Ar/O ratio.

**2.2. Argon abundance from solar flares**

Feldman & Widing (1990) measured abundances for an impulsive flare believed to be close to the solar photospheric composition of the high FIP elements. The conclusion that this flare is near photospheric in composition was based on the O/Mg ratio, which appeared consistent with the older photospheric O/Mg ratio. This conclusion may no longer hold because the new solar O abundance is less. The difference between the solar and impulsive flare O/Mg ratios then suggest that O and Mg in the flare were fractionated from their photospheric values.

Feldman & Widing (1990) give N(Ar)/N(Mg) = 0.15$\pm$0.05 and N(O)/N(Mg) = 24$\pm$4 for the impulsive flare. Feldman & Laming (2000) revised this to N(Ar)/N(Mg) = 0.13$\pm$0.05. Feldman & Widing (2003) further revised it to N(Ar)/N(Mg) = 0.11($\pm$0.05), which is used here to calculate the solar Ar abundance. This ratio combined with N(Mg) = 1.02$\times 10^6$ (Lodders 2003) gives N(Ar) = (1.12$\pm$0.51)$\times 10^5$ or A(Ar) = 6.59$\pm$0.16 for the photosphere. Pottasch & Bernard-Salas (2006) used a similar approach to obtain the Ar abundances from the Ar/Mg ratio. As mentioned above, FIP fractionation of O and Mg is a likely possibility. Therefore the Ar abundance is calculated from the Ar/O ratio, because O has a closer FIP (13.6 eV) to that of Ar. From the Ar/Mg and O/Mg ratios above, we find $N(Ar)/N(O) = (4.58\pm2.10)\times10^{-3}$. This ratio combined with $A(O) = 8.69\pm0.05$ from Allende-Prieto et al. (2001) suggests a photospheric abundance of $A(Ar) = 6.35\pm0.16$ or $N(Ar) = 64,700\pm30,000$. This is 0.24 dex less than the Ar abundance from the Ar/Mg ratio, and may reflect the size of the FIP fractionation between Mg and O. However, the large uncertainty of ~50% for this ratio does not make comparisons too meaningful. Here the Ar abundance from the $N(Ar)/N(O)$ ratio is preferred because any FIP related fractionations should be smaller for Ar and O than for Ar and Mg.

Finally, a direct determination of the solar flare abundance $N(Ar)/N(H) = (2.8\pm0.2)\times10^{-6}$ by Phillips et al. (2003) suggests $A(Ar) = 6.45\pm0.03$ for the solar Ar abundance, if no FIP fractionations between H (13.6 eV) and Ar (15.8 eV) apply.

**2.3 Argon abundance from SEP**

Grevesse & Sauval (1998) compared the derived photospheric Ar value of Young et al. (1997) to the SEP abundance of Ar by Reames (1998) and selected the latter as $A(Ar) = 6.40\pm0.06$ because of the smaller uncertainty. The Ar abundance, $A(Ar) = 6.39\pm0.03$, given in Reames (1998) is derived from the SEP Ar/O ratio of 0.0033 and an adopted O abundance of $A(O)=8.87$. Asplund et al. (2005) use a lower photospheric abundance of $A(O) = 8.66$, which in combination with the Ar/O ratio by Reames (1998) gives their recommended $A(Ar) = 6.18$. The abundance $A(O)=8.69$ from Allende-Prieto et al. (2001) used here suggests $A(Ar) = 6.21$, which is given in Table 1.

However, the Ar abundance listed by Reames (1998) may only give a lower limit to the photospheric Ar abundance. The FIP correlated elemental fractionations already mentioned for the solar wind above also occur for SEPs. Elements with high-FIP are less abundant than low-FIP elements in SEP relative to photospheric abundances (see Fig. 2 of Reames 1998). Anders & Grevesse (1989) also considered SW and SEP data to derive their photospheric Ar abundance, but not before adding a factor of 0.65 dex to the coronal values to correct for coronal/photospheric fractionations. However, the fractionation of high-FIP elements from low-FIP elements seems to be somewhat larger in SEPs than in the SW (e.g., Meyer 1993, Feldman & Laming 2000), so a common correction factor for FIP related element fractionations in SW and SEP may not apply. Furthermore, with the revised solar abundances of many high-FIP elements, the solar-normalized abundance trends in the different solar corpuscular sources need to be re-evaluated. It is not clear yet if the high-FIP elements fractionate from low FIP elements by a constant factor above a certain threshold, or if the elemental fractionations relative to solar are a steady function of FIP (or some other variables such as first ionization time or charge-to-mass ratios). Only then it would be possible to better correct the observed Ar/element abundance ratios in SW and SEPs for the fractionations from photospheric values.

**3. Ar abundance constraints from noble gas measurements in Jupiter**

Like the other gas-giant planets in our solar system, Jupiter must have captured a large amount of gas from the solar nebula because this is necessary to explain its H and He content. The noble gases were likely captured along when solar nebula gas accreted to Jupiter. Thus, the solar system Ar abundance can be estimated from the Galileo probe measurements of the noble gas abundances in Jupiter's atmosphere by multiplying the measured Ar/Kr and Ar/Xe ratios by the solar Kr and Xe abundances, respectively. Mahaffy et al. (2000) report mixing ratios of $Ar/H_2 = (1.82\pm0.36)\times10^{-5}$, $Kr/H_2 = (9.3\pm1.7)\times10^{-9}$ and $Xe/H_2 = (8.9\pm1.7)\times10^{-10}$. These ratios have an uncertainty of about 20%, which is kept for the Ar abundance estimate below. Using the solar abundances of $N(Kr) = 55.15$ and $N(Xe) = 5.391$ (Lodders 2003) and the Jovian $N(Ar)/N(Kr) = 1,960\pm390$ and $N(Ar)/N(Xe) = 20,450\pm4,090$ from the data by Mahaffy et al. (2000), the solar Ar abundance is $N(Ar) = (1.09\pm0.22)\times10^5$ (from Kr) and $(1.10\pm0.22)\times10^5$ (from Xe). This corresponds to a photospheric value of $A(Ar)_{photo} = \log N(Ar) + 1.540 = 6.58\pm0.08$, and a protosolar value of $A(Ar)_{solsys} = \log N(Ar) + 1.614 = 6.66\pm0.08$.

This estimate for the solar Ar abundance is straightforward if the noble gases in Jupiter never fractionated from their original solar ratios. This estimate assumes that Ar, Kr, and Xe accreted to Jupiter in solar proportions, and that they remained so in Jupiter's observable atmosphere without being fractionated by interior differentiation processes.

The $Kr/H_2$ and $Xe/H_2$ ratios on Jupiter relative to the respective solar system ratios are $2.05\pm0.37$ (Kr) and $2.00\pm0.38$ (Xe), indicating that Kr and Xe are equally fractionated from the solar $Kr/H_2$ and $Xe/H_2$ ratios. This fractionation can be done either by adding Kr and Xe or by reducing the amount of hydrogen in the observable atmosphere. However, Kr and Xe are not fractionated from each other; and in particular, the observed Jovian

Kr/Xe ratio is solar. This is significant because of the hypothesis that noble gases were delivered to Jupiter by planetesimals containing noble gases as clathrates, adsorbed on solids or trapped in ices (e.g., Owen et al. 1999, Gautier et al. 2001). All these processes favor sequestering heavy noble gases over lighter ones (e.g., Anders 1963, Lodders 2004). If any significant amounts of noble gases were delivered by planetesimals to Jupiter, the Jovian Ar/Kr, Ar/Xe, and Kr/Xe ratios would be lower than the respective solar ratios because planetesimals would preferentially bring Xe, and to a lesser extent also Kr. In that case, the calculated solar Ar abundance given above would be underestimated. However, the solar Kr/Xe ratio on Jupiter suggests no fractionated noble gas delivery.

Interior differentiation can lower the H, He, and Ne abundances in the observable Jovian atmosphere below solar by formation of a H-metallic liquid and core (see, e.g., Lodders 2004). This would lead to $Kr/H_2$ and $Xe/H_2$ larger than solar. This could have affected the relative Ar, Kr, and Xe abundances. Again, the observed solar Kr/Xe ratio in Jupiter suggests no fractionations between elements. Overall, currently there are no compelling arguments why the observed Jovian Ar/Kr and Ar/Xe ratios should not be representative of the solar abundance ratios.

**4. Argon abundance from the nuclear semi-equilibrium abundance method**

The Ar abundance in Lodders (2003) was obtained from an interpolation of the $^{36}$Ar abundance between $^{28}$Si and $^{40}$Ca, where local nuclear statistical equilibrium is well-established (Cameron 1973, 1982). The $^{36}$Ar abundance derived from this semi-equilibrium abundance method was scaled to the elemental photospheric Ar abundance of A(Ar) = 6.55±0.08 using the isotopic abundances listed in Table 6 of Lodders (2003). This abundance was assigned an uncertainty of 20%.

**5. Argon abundances in B stars**

Elemental abundances in early B stars in the solar neighborhood are known to be close to those in the sun (e.g., Cunha & Lambert 1992, Daflon et al. 2004.). Argon abundances can be spectroscopically determined in the photospheres of these hot stars. Kennan et al. (1990) and Holmgren et al. (1990) derived A(Ar) = 6.49±0.1 (LTE analyses) and 6.50±0.05 (non-LTE-analyses) for a sample of B stars, respectively. While this paper was in revision, Lanz et al. (2007) submitted a paper with non-LTE abundance analyses for 10

B-stars in the Orion association. They derive an average A(Ar) = 6.66±0.06, which is preferred over the older abundance analyses.

The abundances in B-stars should be compared to the proto-solar abundances and not necessarily to solar photospheric abundances. Element settling in the Sun has reduced the protosolar abundances by 0.074 dex for elements heavier than He over the past 4.6 Ga (abundance are related A(Ar)$_{solsys}$ = log N(Ar) +1.614, or A(Ar)$_{solsys}$ = A(Ar)$_{photo}$ +0.074, see Lodders 2003). In contrast, element settling in the younger B stars should not have altered their photospheric abundances by much. If this effect is taken into account, the solar photospheric Ar abundance from B stars is 6.59±0.06 (Table 1). However, within uncertainties, the abundances of other elements in B-stars are consistent with either proto-solar or photospheric abundances (e.g., Lyubimkov et al. 2005).

**6. Argon abundances in HII regions**

Elemental abundances including Ar were determined in the nearby Orion nebula and in the HII region NGC 3576 of similar metallicity. Esteban et al. (2004) found A(Ar) = 6.62±0.05 for the Orion nebula and Garcia-Rojas et al. (2004) found A(Ar) 6.61± 0.08 for NGC 3576. Other elements appear to be about 0.10-0.20 dex higher in abundance than in the solar photosphere (irrespective of whether photospheric abundances from Lodders (2003) or Asplund et al. (2005) are used), indicating that both the Orion nebula and NGC 3576 are of slightly higher metallicity. Assuming that Ar is similarly enriched by 0.1 to 0.2 dex in the HII regions through galactic chemical evolution since formation of the solar system, and/or that heavy elements settled from the solar photosphere, suggests an Ar abundance of A(Ar) = 6.47±0.10 (6.52 – 6.42) for the solar photosphere.

Esteban et al. (2004) and Garcia-Rojas et al. (2004) compare the Ar abundances of these HII regions to the photospheric Ar abundance from Asplund et al. (2005). This suggests that Ar in the Orion nebula is 0.44 dex higher than solar (Table 15 of Esteban et al. 2004) and is 0.41dex higher in NGC 3576 (Table 14 of Garcia-Rojas et al. 2004). This is considerably different from the 0.10-0.20 dex enrichment for other elements. There are no reasons why Ar should be more abundant than other heavy elements in these H II regions, which suggests that the adopted solar Ar abundance by Asplund et al. (2005) is too low.

## 7. Argon abundance from planetary nebulae

Estimates for the proto-solar Ar abundance can be obtained from Ar abundances in planetary nebulae (PN) because the Ar abundance is not expected to be significantly affected by nucleosynthesis during the asymptotic giant branch stage of their progenitor stars. Henry et al. (2004; henceforth H04) obtained Ar abundances in PN as a function of galactocentric radius and found $\log(Ar/H)+12 = 6.58\pm0.079 - 0.030\pm0.010$ R(kpc). In the literature, the distance of the Sun from the galactic center is taken as either 8.0 or 8.5 kpc, which give indistinguishable abundances of 6.34±0.16 and 6.33±0.16, respectively. Pottasch & Bernard-Salas (2006; henceforth PB06) give another data set which can be fitted to $\log(Ar/H)+12 = 7.30\pm0.19 - 0.11\pm0.03$ R(kpc). This suggests A(Ar) = 6.45±0.41 at 8 kpc and 6.39±0.42 at 8.5 kpc.

Direct averages for PN within $7.5 \leq R \leq 9.5$ kpc, around the solar galactocentric radius taken as 8.5 kpc here, give A(Ar) = 6.48(+0.19/-0.34) for 30 PN from H04 and 6.46(+0.16/-0.26) for 9 PN by PB06. These nearly identical averages are probably more reliable than the abundances from the fits, but still have relatively large uncertainties.

The data set by PB06 is relatively small and excluding the highest and lowest data point gives an average with lower uncertainties. The overall low metallicity of Hu 1-2 may justify to exclude its value from this dataset, and without it, the average is 6.49 (+0.17/-0.21). Also excluding the highest value for BD+30 3639, gives an average 6.44 (+0.12/-0.16) for of the remaining 7 data points. However, there is no obvious reason to exclude BD+30 3639 which has comparable metallicity from O and S abundances as the other objects in the 7.5-9.5kpc range.

The average of all data by H04 in the 7.5-9.5kpc range also has a relatively large uncertainty. Removing the objects with the lowest (IC 418) and largest (NGC 2792, NGC3211) values gives an average of 6.45 (+0.17/-0.29) for 27 objects. This average is not much improved and reflects the spread in the existing data. Typical uncertainties quoted for the individual Ar abundances by H04 are ~15%, and reach up to 60% in some cases. PB06 note that Ar abundance determinations could be 30% uncertain. In addition, the statistics may be further complicated by the fact that different galactocentric distances for the sun are adopted in the different studies, which affects the distance scales for the PN and therefore which objects are within the 7.5-9.5 kpc sample. Hence, it does not make sense to alter the statistical samples for apparent outliers, and the averages of all PN in the 7.5-9.5 kpc range are used here. An overall proto-solar abundance of A(Ar) = 6.47±0.34 from PN is adopted in Table 1.

This Ar value from PN should be representative for the proto-solar abundance, and the solar photospheric abundance should be 0.074 dex lower, a similar situation as for B-stars in §5. Considering the relatively large uncertainty in the PN Ar abundance, applying such a small correction for converting proto-solar to photospheric abundances is not very meaningful, but is done for consistency in Table 1.

## 8. Recommended Argon Abundance and Summary

Table 1 summarizes the data discussed above and Figure 1 gives a graphical comparison. The first three values in Table 1 are the photospheric Ar abundances from three different compilations. The solar corpuscular radiation sources (§2) are give an average of A(Ar) = 6.37±0.13 for photospheric Ar (Table 1). The Ar abundance from the impulsive flare Ar/Mg ratio (labeled IFMg in Figure 1) was not included in this average for reasons discussed in §2. The Ar abundance estimates from noble gas measurements in Jupiter (§3), abundances in B stars (§5) and HII regions (§6), and abundances in planetary nebulae (§7) represent proto-solar abundance equivalents (open symbols) and are scaled to photospheric metallicity (closed symbols) where appropriate. The (unweighted) grand average for the photospheric Ar abundance from all sources and the value from the nuclear semi-equilibrium method (§4) gives A(Ar) = 6.50±0.10.

The recommended solar photospheric argon abundance is A(Ar) = 6.50±0.10. This corresponds to an Ar abundance of N(Ar) = 91,200±23,700 on the cosmochemical abundance scale. The protosolar or solar system abundance, which takes into account the heavy element settling from the photosphere, is A(Ar)$_{solsys}$ = 6.57±0.10.

*Acknowledgements*: I thank Nick Sterling for his interest in the solar Ar abundance, which was an incentive to write up these results. I thank the referee for thoughtful comments. This work was supported in part by NASA grant NNG06GC26G and the McDonnell Center for the Space Sciences, Saint Louis.

TABLE I. ARGON ABUNDANCES

| | A(Ar) | References |
|---|---|---|
| Sun Photosphere | 6.56±0.10 | (1) |
| Sun Photosphere | 6.55±0.08 | (2), §4 |
| Sun Photosphere | 6.18±0.08 | (3) |
| | | |
| Solar wind in meteoritic metal | 6.49±0.03 | (4) |
| Solar wind, mainly lunar soil, from Ar/O | 6.29±0.10 | (5) |
| Impulsive flare, from Ar/Mg | 6.66±0.14 | (6,7) |
| Impulsive flare, from Ar/O | 6.42±0.11 | (6,7) |
| Solar flare, from Ar/H | 6.45±0.03 | (8) |
| Solar energetic particles | 6.21±0.03 | (9) |
| Average solar corpuscular radiation [a] | 6.37±0.13 | |
| | | |
| Protosolar from Jovian Ar/Kr, Ar/Xe | 6.66±0.08 | See §3 |
| Photospheric from Jovian Ar/Kr, Ar/Xe | 6.58±0.08 | See §3 |
| | | |
| B stars (= proto-solar) | 6.50±0.10 | (10,11) |
| B-stars (= proto-solar) | 6.66±0.06 | (12) §5 |
| Photospheric from B stars | 6.59±0.06 | (12) §5 |
| HII regions (= proto-solar) | 6.62±0.08 | (13,14) |
| HII regions (scaled to photospheric metallicity) | 6.47±0.10 | §6 |
| Planetary nebulae (= proto-solar) | 6.47±0.34 | (15) §7 |
| Planetary nebulae (= proto-solar) | 6.46±0.26 | (16) §7 |
| Photospheric from PN | 6.40±0.34 | (15) §7 |
| Grand average [b] | 6.50±0.10 | §8 |

REFERENCES: (1) Anders & Grevesse 1989; (2) Lodders 2003; (3) Asplund et al. 2005; (4) Murer et al. 1997; (5) Young et al. 1997; (6) Feldman & Widing 1990; (7) Feldman & Laming 2000; (8) Phillips et al. 2003; (9) Reames 1998; (10) Keenan et al. 1990; (11) Holmgren et al. 1990; (12) Lanz et al. 2007, (13) Esteban et al. 2004; (14) Garcia-Rojas et al. 2004; (15) Henry et al. 2004; (16) Pottasch & Bernard-Salas 2006.

Notes: [a] Ar from impulsive flare Ar/Mg ratio is not included in average

[b] Unweighted grand average (± 1σ) from "photospheric" Jupiter, the average of solar corpuscular radiation sources, B stars (12), HII regions, and PN (all scaled to photospheric metallicity), and the nuclear semi-equilibrium value from (2).

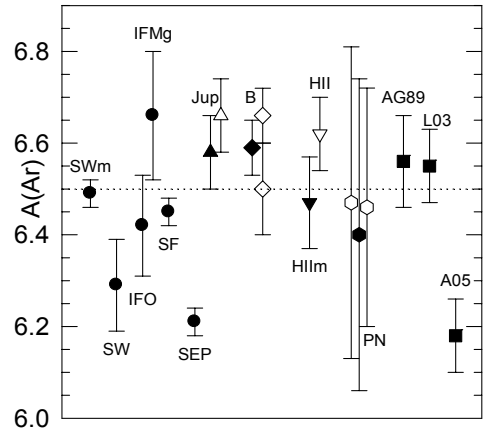

Figure 1. Solar argon abundance estimates from sources listed in Table 1. Circles indicate solar sources (§2), upside triangles are abundances from Jupiter (§3), diamonds show average abundances in B stars (§5), HII region abundances are shown as down triangles (§6) and hexagons are for planetary nebulae (§7). Closed symbols are for values that correspond to photospheric Ar abundances. Open symbols refer to proto-solar abundance equivalents. Squares show the previously selected photospheric Ar abundance by Anders & Grevesse (1989), Lodders (2003), and Asplund et al. (2005). The dashed line is the Ar abundance A(Ar) = 6.50±0.10 recommended here (§8).